\documentclass{webofc}
\usepackage[varg]{txfonts}   
\usepackage{comment}
\usepackage[breaklinks, citecolor=blue, linkcolor=blue, colorlinks=true, debug, baseurl=' ']{hyperref}

\begin{document}
\title{Observations with KIDs Interferometer Spectrum Survey (KISS)}
%
%

\author{A.~Fasano \inst{\ref{LAM}}\thanks{Corresponding author: Alessandro Fasano, \url{alessandro.fasano@lam.fr}}
	\and A.~Catalano\inst{\ref{LPSC}}
	\and J.~F.~Mac\'ias-P\'erez\inst{\ref{LPSC}}
	\and M.~Aguiar \inst{\ref{IAC},\ref{laguna}}
	\and A.~Beelen\inst{\ref{LAM}}
	\and A.~Benoit \inst{\ref{Neel}}
	\and A.~Bideaud \inst{\ref{Neel}}
	\and J.~Bounmy\inst{\ref{LPSC}}
	\and O.~Bourrion \inst{\ref{LPSC}}
	\and G.~Bres \inst{\ref{Neel}}
	\and M.~Calvo \inst{\ref{Neel}}
	\and J.~A.~Castro-Almaz\'an \inst{\ref{IAC},\ref{laguna}}
	\and P.~de~Bernardis\inst{\ref{Roma}}
	\and M.~De~Petris\inst{\ref{Roma}}
	\and A.~P.~de~Taoro \inst{\ref{IAC},\ref{laguna}}
	\and M.~Fern\'andez-Torreiro\inst{\ref{IAC},\ref{laguna}}
	\and G.~Garde \inst{\ref{Neel}}
    \and R.~G\'enova-Santos \inst{\ref{IAC},\ref{laguna}}
	\and A.~Gomez \inst{\ref{madrid}}
	\and M.~F. G\'omez-Renasco \inst{\ref{IAC},\ref{laguna}}
	\and J.~Goupy \inst{\ref{Neel}}
	\and C.~Hoarau\inst{\ref{LPSC}}
	\and R.~Hoyland \inst{\ref{IAC},\ref{laguna}} 
	\and G.~Lagache\inst{\ref{LAM}}
	\and J.~Marpaud\inst{\ref{LPSC}}
	\and M.~Marton\inst{\ref{LPSC}}
	\and A.~Monfardini \inst{\ref{Neel}}
	\and M.~W.~Peel \inst{\ref{IAC},\ref{laguna}}
	\and G.~Pisano \inst{\ref{Cardiff}}  
	\and N.~Ponthieu\inst{\ref{IPAG}}
	\and R.~Rebolo \inst{\ref{IAC},\ref{laguna}}
	\and S.~Roudier\inst{\ref{LPSC}}
	\and J.~A.~Rubi\~no-Mart\'in\inst{\ref{IAC},\ref{laguna}}
	\and D.~Tourres\inst{\ref{LPSC}} 
	\and C.~Tucker \inst{\ref{Cardiff}}
	\and C.~Vescovi \inst{\ref{LPSC}}
}

\institute{
    Aix Marseille Univ, CNRS, CNES, LAM, Marseille, France 	\label{LAM}
	\and
    Univ. Grenoble Alpes, CNRS, LPSC/IN2P3, 38000 Grenoble, France
	\label{LPSC}
	\and
	Univ. Grenoble Alpes, CNRS, Grenoble INP, Institut Néel, 38000 Grenoble, France
	\label{Neel}
	\and
	Instituto de Astrofísica de Canarias, C/Vía Láctea, E-38205 La Laguna, Tenerife, Spain
	\label{IAC}
	\and
	Universidad de La Laguna, Dept. Astrofísica, E-38206 La Laguna, Tenerife, Spain
	\label{laguna}
	\and
	Dipartimento di Fisica, Sapienza Universit\`a di Roma, Piazzale Aldo Moro 5, I-00185 Roma, Italy
	\label{Roma}
	\and
    Centro de Astrobiología (CSIC-INTA), Torrejón de Ardoz, E-28850, Madrid, Spain 
	\label{madrid}
	\and
	Astronomy Instrumentation Group, University of Cardiff, The Parade CF24 3AA, UK
	\label{Cardiff}
	\and
	Univ. Grenoble Alpes, CNRS, IPAG, 38400 Saint Martin d'Hères, France 
	\label{IPAG}
}

\abstract{
We describe the preliminary on-sky results of the KIDs Interferometer Spectrum Survey (KISS), a spectral imager with a 1\,deg field of view (FoV). The instrument operates in the range 120--180\,GHz from the 2.25\,m \mbox{Q-U-I JOint TEnerife telescope} in Teide Observatory (Tenerife, Canary Islands), at 2\,395\,m altitude above sea level. Spectra at low resolution, up to 1.45\,GHz, are obtained using a fast (3.72\,Hz mechanical frequency) Fourier transform spectrometer, coupled to a continuous dilution cryostat with a stabilized temperature of 170\,mK that hosts two 316-pixel arrays of lumped-element kinetic inductance detectors. KISS generates more than 3\,000 spectra per second during observations and represents a pathfinder to demonstrate the potential for spectral mapping with large FoV.
We give an overall description of the spectral mapping paradigm and we present recent results from observations, in this paper.
}

\maketitle

\section{Introduction}
\label{intro}

In the millimeter (mm) astronomy domain, there is a strong demand from the scientific community to develop multi-band instruments for component separation, foreground characterization, cosmic microwave background spectral distortions and line intensity mapping \cite{esa}. An interesting instrumental candidate to meet these demands is the exploitation of Fourier transform spectrometers (FTSs) and, in particular, the Martin-Puplett interferometer (MPI) configuration \cite{mpi,kiss-ltd}. The case of ground-based experiments has an additional requirement: the atmospheric fluctuations have to be addressed. For this purpose, FTSs have to be coupled with fast detectors, and kinetic inductance detectors (KIDs) are the fastest available in large format arrays at mm-wavelength \cite{masi,catalano}.

In this scientific framework, we have developed the KIDs Interferometer Spectrum Survey (KISS), which uses two arrays of KIDs coupled to an MPI. The reference source of such a differential FTS configuration can be set between the image of the de-focused sky and a cryogenics stage, see \cite{kiss-ltd} for more details.
KISS allows us to exploit a wide instantaneous field of view (1\,deg) and a spectral resolution as fine as 1.45\,GHz in the 120--180\,GHz frequency band. The instrument is installed on the 2.25-meter Q-U-I JOint TEnerife (QUIJOTE) telescope at the Teide Observatory, in Tenerife and has been operational since February 2019.
KISS is also the pathfinder of the new CarbON CII line in the post-rEionization and ReionizaTiOn epoch project (CONCERTO) \cite{concerto-aa}.

In this paper, we give an overall description of the spectral mapping paradigm and we present recent results from the last year of observations.
In Section~\ref{sec:mapping}, we describe the necessity for fast scanning strategy and we explain the spectral mapping technique.
Section~\ref{sec:onsky} presents the preliminary on-sky results with point sources in spectral mapping reduction and we address the data reduction of the KISS project.

\section{Spectral mapping with FTS fast scanning}
\label{sec:mapping}

Ground-based experiments have to deal with a particular additional issue with respect to space-borne telescopes. Despite the ease to debug and modify the hardware of telescopes on the ground, even after the deployment, an additional source of contamination is present: the atmosphere, both in terms of absorption and emission.
This limits the frequency range over which they can be used. The emission has two main effects: 1) it increases the background (photon) noise with respect to what is expected for satellite experiments, and 2) leads to low-frequency noise induced by low-frequency drifts in the atmospheric emission. For the first, it requires an increase of the necessary integration time for a given signal-to-noise ratio. However, this requirement is compensated by the lower cost of a ground-based facility.  For reducing the latter, there are specific observation strategies for different instruments that prevent the introduction of systematic errors in the signal. This problem has to be carefully addressed to avoid affecting the accuracy of the calibration. In the case of FTS, the solution is to acquire the full interference pattern, the so-called ``interferogram'', on times scales shorter than those of the atmospheric fluctuations.

\begin{figure}[h]
\centering
\includegraphics[scale=0.3]{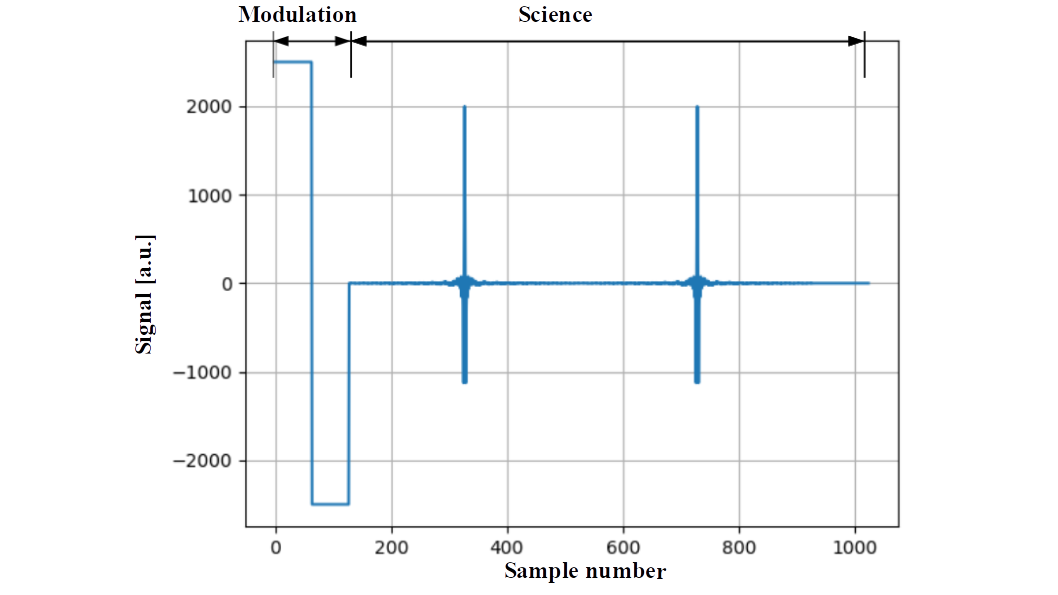}
\caption{KISS data block recorded at 3.72\,Hz: frequency signal as a function of the sample number. There are a total of 1024 data samples: the first 128 are dedicated to the modulation and the remaining ones are to the forward and backward interferogram acquisition. The signal is the result of the median value subtraction (i.e., by scaling to 0\,Hz the median value), which represents the photometry, i.e., the signal integrated over the electromagnetic bandwidth.}
\label{fig:modulation}
\end{figure}

State-of-the-art detectors in the mm domain reach a white noise level dominated component is the photon noise, which is the intrinsic noise derived by the random nature of the incoming photons. The best noise performance is matched if this condition is verified and the detectors are named photon-noise limited. The second higher noise component is produced predominantly by the electronic noise. In addition, there is the noise coming from the state fluctuation of the superconducting particles, called generation-recombination noise, in the KID case \cite{kid}.
The slow fluctuations of the atmosphere are translated to a 1/f noise in the frequency domain, whose intersection with the white noise forms a knee shape. The frequency of the intersection, the knee frequency, is particularly important for ground-based experiments. The recording of every single interferogram at stable atmospheric absorption (in other words, faster than the 1/f knee) is a prior requirement, in the specific KISS case. We use the so-called ``fast scanning'' technique for interferometric pattern sampling: we record the interferogram on the fly while continuously oscillating the roof mirror, which introduces the optical path difference (OPD) while pointing the telescope on-sky (see \cite{casper2} for a description of the phase and amplitude modulation alternative techniques, not suitable for our utilization).
The typical 1/f atmospheric knee is expected to be at $\sim$1\,Hz \cite{ritacco}, in the heritage of the New IRAM KID Array 2 (NIKA2) \cite{nika2} at the Institut de Radioastronomie Millimétrique (IRAM) telescope in Pico Veleta (Spain).
This, coupled with the requirement on the spectroscopic resolution, leads to the acquisition of hundreds to thousands of samples on time scales of 0.25 seconds. To achieve this performance and ensure a large field of view (FoV), we need large arrays of fast detectors and readout as is the case for the KIDs.
To meet these requirements KISS integrates two full interferograms at 3.72\,Hz (0.27\,s).
The presence of a double interferogram is due to the adopted sampling technique of the interferogram (see \cite{spencer} for a detailed discussion): we double-sample the interferogram by measuring slightly above the zero optical path difference (ZPD) point that corresponds to the null interference. The first and second interferograms are the result of the forward and backward mirror motion, respectively, as shown in Fig.~\ref{fig:modulation}.

In addition, we have introduced an electrical modulation before each interferogram recording, as shown in Fig.~\ref{fig:modulation}. This readout modulation is adopted to improve the sky signal reconstruction accuracy for types of instruments for which a fast sampling frequency is required, both to remove atmospheric fluctuations and to perform full spectroscopic measurements on each sampled sky position.
Kinetic inductance detectors are superconducting resonators that are read out by injecting a reference signal in the circuit and recording the output one. The physical quantity that can be directly translated into an astrophysical signal is the resonance frequency, which is different for each single KID. The optical load changes this electrical parameter and by recording it we retrieve the astrophysical signal. The raw data from the KID read-out system is the complex signal $(I,Q)$ from which we can compute the phase signal $\left( \phi = \arctan2 ( I/Q ) \right)$. Before each on-sky scan, the single KID is optimized (tuned) for the background at the observation time.
The electrical modulation introduced before each interferogram sampling allow us to follow the background evolution, reducing the downsides from the evolved optical load that introduces imperfection on the initial tuning approximation. In addition, it allows us to convert the raw $\phi$ to the resonance frequency: for a detailed discussion see \cite{kiss-aa}.

This whole method is translated into cubes of data composed of the x-y positions on the sky plus the electromagnetic information: we retrieve a whole interferogram per sky position.
Starting from the raw interferogram data presented above we go to maps of the sky as follows: 1) using the modulation we calibrate the signal in frequency shift, 2) the data corresponding to the forward and backward interferograms are separated; 3) the interferograms are cleaned from low-frequency drifts and the ZPD is obtained, by regridding the interferograms on the OPD and identifying the maximum or minimum signal; 4) the interferograms in the time domain are projected into maps using all detectors, and 5) the final spectral maps are obtained from the Fourier transform of the interferograms accounting for the ZPD.
The result is a set of N maps at the N bins of the frequency bandwidth.

In this paper, we present uncalibrated signals, which are to be intended as resonance frequency ones, as previously mentioned.

\section{On-sky results}
\label{sec:onsky}

In this section, we show early results in uncalibrated signals of Jupiter and the calibrated flux of Venus.
In previous works, we have already demonstrated the possibility of exploiting an FTS
as a pure photometer. In particular, observing a bright and wide source like the Moon \cite{kiss-nika2} which acted as a quick-look demonstrator for our challenging instrument, and calibrating Venus with Jupiter observations \cite{kiss-aa}.

\begin{figure}[ht]
	\centering
	\includegraphics[width=.6\textwidth]{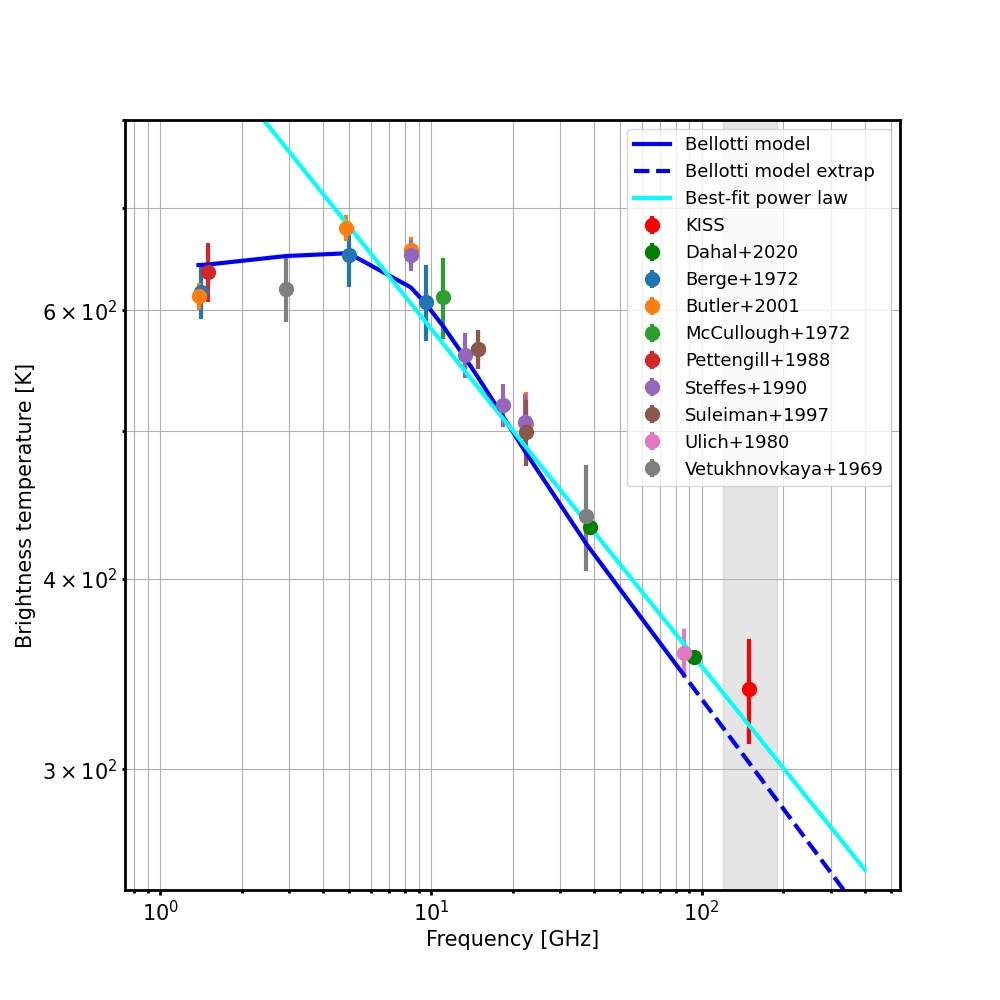}
	\caption{Venus brightness temperature as a function of frequency. We present a collection of radio and mm wavelength measurements including those of KISS. The blue and cyan lines represent two models for the Venus brightness temperature. \cite{kiss-aa}}
	\label{fig:venus}
\end{figure}

We exploited Jupiter to calibrate our detector response in an astrophysical signal and obtain the first scientific result, as shown in Fig.~\ref{fig:venus}. We measure the brightness temperature of Venus at a frequency that has not been measured before. We obtain a brightness temperature of $T^{\rm{KISS}}_{b,\rm{Venus}} = 338 \pm 27$\,K  at 150\,GHz (the central frequency of the 120--180\,GHz KISS band and the incertitude represents the 1-sigma level of confidence). We also show the ``Bellotti'' model \cite{bellotti} as a solid blue line for the Venus brightness temperature and its extrapolation of a simple power-law as a dashed line at mm wavelengths. In solid cyan line, we represent the best-fit power law.  We observe that the Bellotti model underestimates the power law at high frequency, for more details see \cite{kiss-aa}.

The next challenge is to obtain the spectral mapping of the two planets.
The coupling between KISS and the Q-U-I JOint TEnerife (QUIJOTE) telescope results in 7\,arcmin (at the central electromagnetic frequency of 150\,GHz) full width at half maximum (FWHM). Venus and Jupiter have respectively $338 \pm 27$\,K and $174.1 \pm 0.2$\,K \cite{jupiter} in thermodynamic temperatures, and 66\,arcsec and 50\,arcsec maximum intrinsic apparent diameters when they are at the closest approach in the orbit. We are, therefore, affected by a strong dilution factor, which is defined geometrically as $ D = \Omega / 2\pi \sigma^2 $, where $\Omega$ is the source solid angle and $\sigma$ is the sigma of the single-pixel $\sigma = FWHM / 2 \sqrt{ 2 \log(2)  } $. In conclusion, the planet signal is diluted within the beam and it results in few kelvins at the different apparent diameters, in the KISS case. This was a major issue for the early observations because we experienced the absence of bright and large-enough primary calibrators to accurately constrain the pointing model at the beginning of the commissioning phase. Despite these difficulties, we managed to detect and systematically observe these two cited planets.

\begin{figure}[h]
\centering
\includegraphics[scale=0.4]{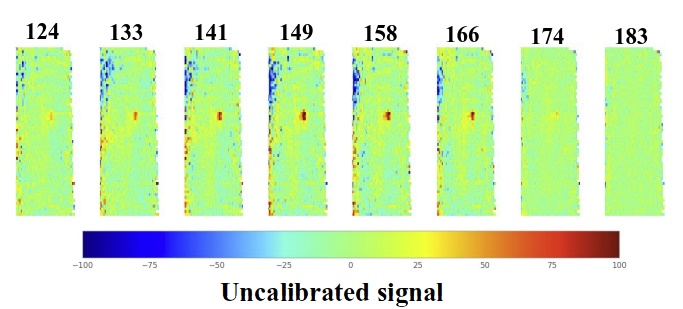}
\caption{Uncalibrated spectral mapping of Jupiter with 8\,GHz resolution in frequency. The large bin has been adopted to increase the sensitivity. We show the mean value from all detectors. The central frequency in GHz is given at the top of each map.}
\label{fig:jupiter_sm}       
\end{figure}

After the first spectral mapping with the Moon presented in \cite{kiss-ltd}, we introduce here the results with spectral mapping capability with point sources with KISS which can be considered as the proper first light of the experiment.
Figure~\ref{fig:jupiter_sm} shows the spectral mapping of Jupiter, projecting the spectra at the given sky position per electromagnetic bin. We observe how, despite the large size of the map needed due to the large FoV, the source is well identified in the center with a signal-to-noise ratio (SNR) of $\sim$10.

The presented spectral maps are the result of a portion of the original (typical) 90x90 arcmin$^2$ sky coverage, obtained in 10\,minutes of total integration time per observation. We can calibrate the spectral response by exploiting the integrated spectrum of such observations: in other words, we obtain a calibration factor between the detector response and the astrophysical signal per electromagnetic bin.

\section{Conclusions}
\label{sec:conclusion}

We have demonstrated the capability of instantaneously mapping large areas of the sky (one-degree diameter FoV) in spectroscopic mode by  exploiting FTS coupled to KIDs.

KISS has played a fundamental role as the test bed for dedicated spectral mapper large-telescope, CONCERTO, which has currently passed its technical commissioning phase, as reported in \cite{concerto-jltp}. 
Despite its pioneering and challenging nature, it has paved the way for the technological advancement of the mm observations ground-based at multi-wavelength large FoV.

In this paper, we have reported the integrated Venus brightness temperature and the uncalibrated spectral maps of Jupiter.\\
\\
\\
\small
{\bfseries \emph{Acknowledgements.}} The KID arrays described in this paper have been produced at the PTA Grenoble microfabrication facility. 
This work has been partially supported by the LabEx FOCUS ANR-11-LABX-0013, the European Research Council (ERC) under the European Union's Horizon 2020 research and innovation program (project CONCERTO, grant agreement No 788212), and the Excellence Initiative of Aix-Marseille University-A*Midex, a French ``Investissements d'Avenir'' program.
This work has been partially funded by the Spanish Ministry of Science under the project AYA2017-84185-P.
Based on observations made with the first QUIJOTE telescope (QT-1), operated on the island of Tenerife in the Spanish Observatorio del Teide of the Instituto de Astrofisica de Canarias.

%
%

\end{document}